# COMMISSIONING OF THE CRYOGENICS OF THE LHC LONG STRAIGHT SECTIONS

A. Perin [1], J. Casas-Cubillos[1], S. Claudet[1], C. Darve[2], G. Ferlin[1], F. Millet[1], C. Parente[1], R. Rabehl[2], M. Soubiran[1], R. van Weelderen[1], U. Wagner[1]

[1] CERN Technology Department,
CH-1211 Geneva 23

[2] Fermi National Accelerator Laboratory
Batavia, Illinois, 60510, USA

**ABSTRACT**

The LHC is made of eight circular arcs interspaced with eight Long Straight Sections (LSS). Most powering interfaces to the LHC are located in these sections where the particle beams are focused and shaped for collision, cleaning and acceleration. The LSSs are constituted of several unique cryogenic devices and systems like electrical feed-boxes, standalone superconducting magnets, superconducting links, RF cavities and final focusing superconducting magnets. This paper presents the cryogenic commissioning and the main results obtained during the first operation of the LHC Long Straight Sections.

## INTRODUCTION

The Large Hadron Collider is made of eight sectors, each made of a circular arc and of two straight sections [1] as is shown in FIGURE 1. The arc parts of the sectors, of an approximate length of 2.8 km include mainly curved 15 m long superconducting dipole magnets for bending the trajectory of the particles along the tunnel shape and superconducting quadrupole magnets (called short straight sections) to focus the beam. Although some special magnets are included at the arc extremities, in the dispersion suppressor areas, most of the fine shaping, cleaning, acceleration, final focusing and collision of the particle beams is performed in straight sections, about 250 m long on each side of the sector called the Long Straight Sections (LSSs). Except for the 60A correctors, all superconducting magnets of the LHC are also powered through cryogenic feedboxes located in the LSSs. While the 22.5 km of arcs are very similar along the circumference of the LHC, the LSS, exhibit in about 4 km a wide range of cryogenic configurations dictated by the specific functions required for controlling the two counter rotating particle beams of the LHC. This paper gives an overview of the cryogenic commissioning of the equipment located in the 8 LSSs of the LHC (The cryogenic distribution line [2] is also present in the LSSs but is not treated in this article).

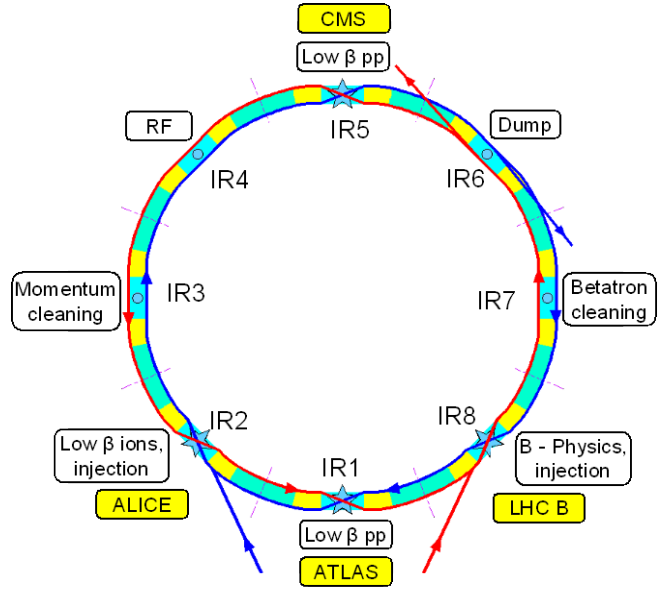

**FIGURE 1.** General layout of the LHC showing the position of the long straight sections.

## CONFIGURATION OF THE LSSs

From the cryogenics supply point of view, the LHC is split into 8 sectors that are cooled independently by 8 cryogenic plants. However as concerns the accelerator optics, and therefore the devices used for beam control in the LSS, there is essentially mirror symmetry around the interaction point (IP) between two sectors. In most case this results in a mirror symmetry also for the cryogenic equipment. It is therefore convenient to refer to the Long Straight Sections by using a name relative to the IP and not the name of the sector (e.g. LSS3R is part of sector 2-3 and located on the right of IP3). The mirror LSS located to the left and to the right on an IP are together referred to as an Interaction Region (IR).

**TABLE 1**. List of main cryogenic devices found in the LSS of LHC.

| Acronym | Description |
|---|---|
| ACS | Beam acceleration cryomodules made of radio-frequency superconducting cavities |
| DFB | Cryogenic electrical feedboxes ( 4 types: DFBA, DFBM, DFBL, DFBX) |
| DSL | Superconducting links (76 m and 517 m long) |
| FFM | Final focusing magnets (quadrupoles or dipoles) |
| SAM | Standalone magnets (single or multiple magnets assemblies) |

**TABLE 2.** List of devices found in the LHC Long Straight sections.

| IR | Main function of IR | Specific cryogenic equipment |
|----|--------------------|------------------------------|
| 1 | ATLAS experiment | 6 FFM, 4 SAM, 6 DFB , 2 DSL (76 m) |
| 2 | ALICE experiment | 8 FFM, 6 SAM, 8 DFB |
| 3 | Beam cleaning | 2 SAM, 5 DFB, 1 DSL (517 m) |
| 4 | Beam acceleration | 6 SAM, 8 DFB, 4 ACS |
| 5 | CMS experiment | 6 FFM, 4 SAM, 6 DFB, 2 DSL (76 m) |
| 6 | Beam dump | 4 SAM, 6 DFB |
| 7 | Beam cleaning | 2 SAM, 4 DFB |
| 8 | LHCb experiment | 8 FFM, 6 SAM, 9 DFB. |

While the adjacent magnets form a continuous cryostat (CC) in the curved parts of the sectors, in the LSSs the cryogenic devices are interspaced with room temperature equipment.

**Cryogenic Devices in the LSSs**

The LSSs of LHC include several types of cryogenic devices used to control the particle beams, accelerate it or power the magnets. The various types of cryogenic devices are listed in TABLE 1. The schematic cryogenic layout of sector 4-5, showing the position of the cryogenic devices is shown in FIGURE 2. The LSSs always start and end with a cryogenic electrical feedbox called DFBA [3,4] that powers the arc section and also provides the mechanical and cryogenic termination to the arc. At the DFBAs the beam tubes emerge from the continuous cryostat of the arc and pass then through superconducting magnets that are installed between stretches of room temperature sections and operate in saturated liquid helium at 4.5 K. These magnets are named sequentially according to their position relative to the IP and are referred to as standalone magnets (SAM). These magnets are powered either by local feedboxes called DFBMs [3,4] (in most LSSs) or remotely, when this first configuration is not possible (only in IP 1 and 5), through superconducting links (DSL) [3,5] of a length of 76 m. The links are themselves powered by special feedboxes: the DFBLs [3,4]. A superconducting link of a length of

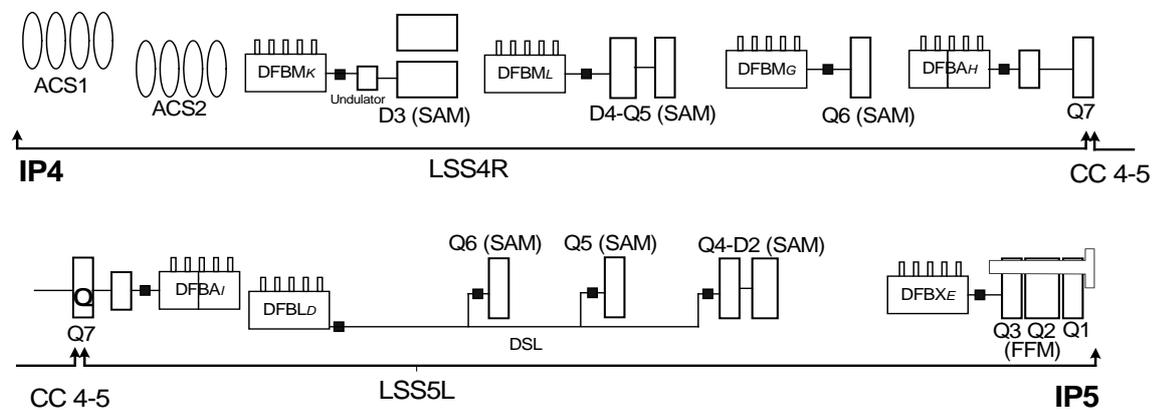

**FIGURE 2.** Schematic configuration of the Long Straight Sections of sector 4-5 of LHC.

517 m is also installed in IR3 where space does not allow the local installation of the power supplies for corrector magnets circuits.

The individual configurations of the LSSs are all different, corresponding to the various functions and experiments of the interaction regions; TABLE 2 summarizes the main functions and experiments of each IP of LHC. In IP1, IP2, IP5 and IP8, an experiment is present and the two beams are made to collide. In order to shape and guide the beams for collision, final focusing magnets (FFMs) [6] are installed on each side of these IPs. The FFMs comprise for each IP three superconducting quadrupoles and a dipole that is superconducting only in IR2 and in IR8. These magnets operate, like the arc magnets, in pressurized superfluid helium at 1.9 K. Being next to where the beams cross, these magnets are subject to large beam induced heat loads, particularly in IP1 and 5 where the byproduct of the high luminosity collisions is a heat load that can reach 200 W for nominal beams. The superconducting FFMs are powered by special feedboxes feedboxes; the DFBXs [6].

The energy is also supplied to the particle beams in two LSSs by superconducting RF cavities installed in IR4. They are grouped into cryo-modules located on each side of IP4 (ACS).

## COMMISSIONING THE LSSs

The cryogenic equipment of the LSSs was commissioned together with the corresponding sectors, closely following the end of the installation. The commissioning started in the first half of 2007 with sector 7-8. The commissioning of the cryogenic system of LHC is described in [7]. After a pause to repair helium leaks in sector 4-5, the bulk of the commissioning operations took place between December 2008 and September 2008. During the second half of August 2008, all LSSs of LHC were at nominal temperature and the green light to inject the beam was given in the first days of September, with the first beams circulating on 10 September 2008. We present here a summary of the commissioning, some of the difficulties that were encountered and the actions that were taken to correct them.

### The Electrical Feedboxes

The cryogenic electrical feedboxes are of two different main types, the ones supplied by CERN [3,4] and the ones supplied by the US-LHC collaboration [6]. None of the feedboxes was tested in cryogenic conditions before their installation in the LHC. The commissioning phase was therefore the first time that the feedboxes were cooled. A more detailed description of their cryogenic commissioning is given in [4] for the CERN supplied ones and in [8] for the US supplied ones.

The LSS include all the cryogenic electrical feedboxes of the LHC, with a total of more than 1300 current leads of which more than 1000 are HTS leads. Each HTS lead and most conventional leads are temperature controlled by individual valves; the nominal operation of the LHC requires therefore the simultaneous operation (and previously the commissioning) of thousands of sensors, data acquisition systems and control loops.

*Global performance*

All DFBs were cryogenically commissioned within the constraints of the global LHC schedule. Except for the main dipole circuits that were powered to 9 kA (nominal 11.8 kA) most circuits were powered to nominal current. After the corrections of diverse difficulties (some of which are presented below) the DFBs performed essentially as expected and, except for two DFBs where non-conformities were present, showed expected heat loads.

*Liquid level measurement*

All vapor-cooled current leads of LHC operate with their cold terminal immersed in saturated liquid helium at a pressure of about 1.3 bar (and a corresponding temperature of 4.5 K). The HTS leads need the liquid helium level to be regulated within 2 cm from the nominal position. This means that the absolute value of the liquid level must be known with a precision of the order of 1 cm. Initially the superconducting level gauges installed in the feedboxes were not calibrated individually. Investigation of the characteristics of the level gauges showed that the required absolute precision could not be achieved with un-calibrated level gauges, as the spread in their properties was found to be too large. A program of calibration was therefore started and all level gauges were calibrated against a geometrical reference in a glass cryostat at CERN central cryogenic laboratory. In addition to the calibration, all liquid level measurements were cross-checked by comparing the boil-off curves with specific geometrical features of the DFBs.

*Water condensation*

During the first commissioning phase it was remarked that water condensation was taking place on the insulating parts at the top of the current leads and on the cryostat next to the leads. Two different approaches were applied to the CERN supplied DFBs and to the US supplied DFBs. For the US-supplied DFBs, a special coating was applied on the cold insulating parts to avoid the creation of a continuous path able to damage the insulation properties [8]. For the CERN-supplied DFBs, the problem was solved with two measures: the temperature of the top part of the current leads was increased to 30°C and copper bridges (in the form of multiple flexible sheets) were added to bring heat to the insulating flange (see FIGURE 3). To further limit the risk of water condensation, the insulating parts and the top of the DFB cryostat were encapsulated into a dry air flow (see FIGURE 3) with a dew point of -45°C tapped from the compressed air system of the LHC. The corrective actions proved to be effective as condensation was consistently suppressed on all leads.

*Current lead control valves*

The helium flow through gas cooled current leads is controlled by proportional solenoid valves, one per lead in most cases. These valves have the advantage of being very compact and do not require electronic components that would be incompatible with the radiation environment. Several sizes of valves are used in the LHC, the smaller ones having a capacity of 0.07 g/s for a differential pressure of 50 mbar. During the initial phases of commissioning several valves, in particular the smaller ones stopped functioning normally and showed a dramatically increased hysteresis. The problem was traced to the presence of small particles in the moving parts of the valves and to the extreme sensitivity of the valves to even particles smaller than 0.1 mm. In the worst cases up to 50 % of the valves of a DFB failed. The particles appear to come from the cryogenic supply line that is also the return header for the helium out of the LHC magnets, a similar contamination has been found in the filters of the cryogenic system. During the commissioning period, the only cure was to replace the valves. Following this experience, all valves have been equipped with 20 μm filters. A program to also clean the manifolds between the filters and the valves is under way.

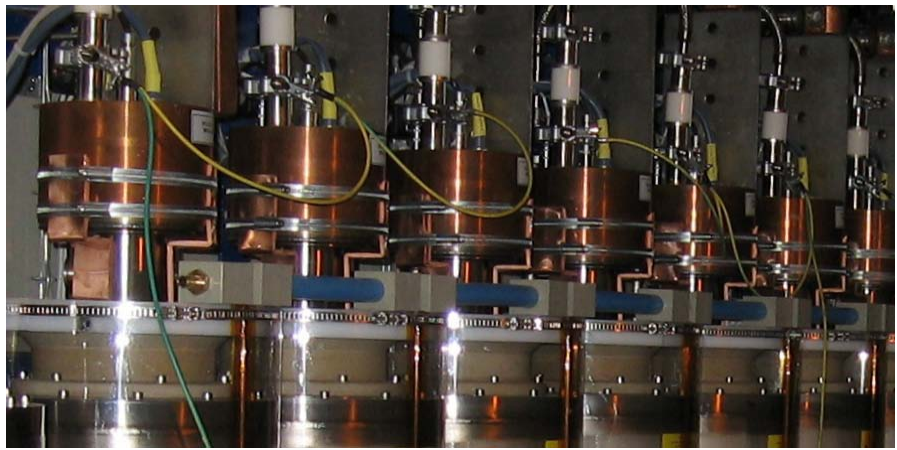

**FIGURE 3.** View of the top part of current leads and of the DFB cryostat showing the copper thermal bridges and the dry air encapsulation system.

**The Superconducting Links**

The usage of superconducting links was a world first for powering accelerator magnets. The commissioning of the four 76 m long superconducting links located in IR1 and in IR5 was performed without encountering specific problems [5]. Their performance was very close to the estimated values and they became rapidly operational. As the links are located at the most distant place from the refrigerators and to ensure a safe temperature margin, helium sub-coolers are installed in the DFBL that feed them. The measured temperatures were as expected and a stability of ± 0.1 K around the nominal temperature was achieved, well within the operational margins of the superconducting cables.

The 517 m long superconducting link located in IR3 was only partially commissioned as an excessive localized heat load of more than 10 W was observed. This created a stratification of temperatures that limited the maximum current that the superconducting cables could carry. The problem was caused by incorrectly designed support system in a flexible section of the link. The commissioning period allowed nevertheless the validation of the cryogenic operation of the link, its sub-cooler and its feedboxes. In particular it was possible to demonstrate the possibility of operating in stable conditions during several days, despite the difficulty of the delayed response due to the length of the link.

**The Standalone Magnets**

Contrary to the superconducting magnets of the arcs and of the final focusing regions, the standalone magnets located in the LSS operate in saturated liquid helium at 4.5 K. In order to ensure that the magnet coils are always covered with liquid helium, the liquid level in the magnets must be regulated. The precision of the regulation is in this case not so important (contrary to the DFBs) as long as the coils are covered. During the commissioning, the magnets were first cooled by flowing cold helium. The cool-down proceeded as expected. However, when the magnets were switched to level regulation mode, in several magnets it was impossible to get a reliable reading of the level of the liquid helium. In addition, boil-off measurements were not coherent with the level readings. The cause was rapidly identified as a problem of pressure equilibration between the main magnet volume and the volume where the level gauge is installed. As can be seen in FIGURE 4, the level gauge is installed in a tube, next to the magnet helium container. This tube is connected to the magnet at its bottom to a tube of about 10 mm in diameter while at the top the pressure equilibration is provided by a much smaller tube of 2 mm diameter. The diameter of this tube was identified as not being large enough to avoid excessive pressure drop or clogging by liquid helium, in particular in cases when the pressures in the supply lines were varying. This was further confirmed by the complete absence of this problem in SAM where the top pressure equilibration tubes were of larger diameter.

The problem was first temporarily cured by installing a pressure equilibration pipe at room temperature, but this led to an offset in the level reading. Thanks to an in-situ individual calibration realized with boil-off measurements it was nevertheless possible to achieve a sufficient degree of precision to ensure that the coils were covered and the magnets could be successfully powered to their nominal current. During the shutdown period 2008-2009 all but two top pressure equilibration tubes have been replaced with larger diameter ones.

**The Final Focusing Magnets**

The final focusing magnets are three low-beta superconducting quadrupoles (also called the triplet) and a dipole magnet that is superconducting in IR2 and IR8. The superconducting magnets operate at 1.9 K in pressurized superfluid helium. After having solved several difficulties with the electrical feedboxes, in particular a redesign of the control system of the resistive gas cooled leads [8], most circuits of the FFMs could be powered to their nominal current values.

From the cryogenic point of view, the specificity of these magnets is the very high beam induced heat load that can be seen at 1.9 K in the high luminosity IR1 and IR5: when the beam collide at nominal luminosity, the beam induced heat load, 200 W, on the three FFMs is expected to be nearly the same as in the 2.8 km of the arc section. As described in [8] it was possible to demonstrate during the commissioning that the cryogenic system could extract such power in a reliable and stable way.

The commissioning operations also unveiled some non-conformities in the thermal configuration of the magnets [8] that have been subsequently corrected during the 2008-2009 shutdown period. Some limitations, in particular on the current ramp rates of the resistive vapor cooled leads, persist and will be dealt with during the next re-commissioning campaign.

**Superconducting Radio-Frequency Cavities**

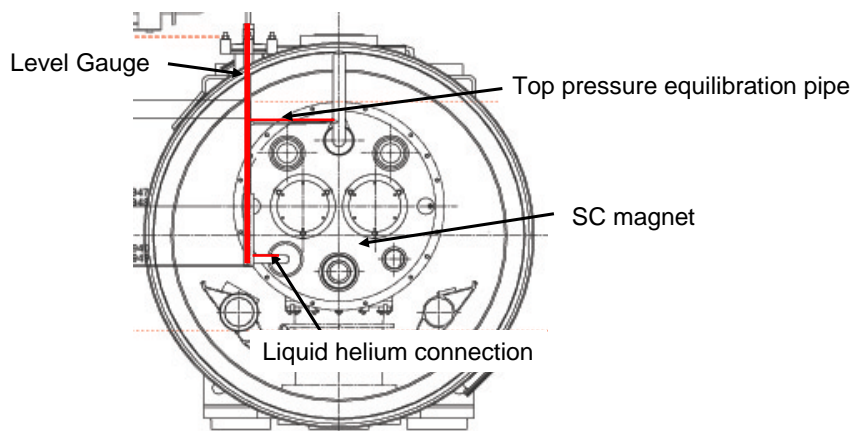

**FIGURE 4.** cross section of a standalone magnet showing the relative positions of the magnet and level gauge and the piping that connects the level gauge to the magnet.

The radio-frequency cavities, located all in IR4 of LHC, operate in saturated liquid helium at 4.5 K. From the cryogenic point of view the ACS are essentially helium tanks, the main requirements being the keep the cavities submerged and to precisely control the pressure of 1.35 bar ± 0.015 bar. No particular problem was encountered during the cryogenic commissioning of the ACS and they operated as expected.

## CONCLUSIONS

The cryogenic commissioning of the very diverse and numerous devices of the long straight sections of the LHC was performed in less than 10 months. For many systems, like the feedboxes and the superconducting links, this was the first time they were cooled. The commissioning phase allowed to start and to tune the operation of 62 electrical feedboxes, more than 1300 current leads, 5 superconducting links, 34 standalone magnet assemblies, 28 final focusing magnets, and 4 cryogenic modules of superconducting cavities. Several problems were found and most of them could be solved during the commissioning period which allowed the first beams to circulate in the LHC.

## ACKNOWLEDGEMENTS

The authors gratefully acknowledge the constant support and the commitment of the operation and commissioning teams of the LHC.## REFERENCES